# Fabrication and physical properties of stable room temperature bulk ferromagnetic graphite

**H. Pardo, R. Faccio and A. W. Mombru**[*]

*Crystallography, Solid State and Materials Laboratory (Cryssmat-Lab), Dequifim, Facultad de Química, Universidad de la República, P.O. Box 1157, Montevideo, URUGUAY*

**F. M. Araujo-Moreira**
*Grupo de Materiais e Dispositivos - CMDMC, Departamento de Física e Engenharia Física, UFSCar, Caixa Postal 676, 13565-905, São Carlos SP, BRAZIL*

**O. F. de Lima**
*Instituto de Física "Gleb Wataghin", UNICAMP, 13083-970, Campinas SP, BRAZIL*

## Abstract

The search for macroscopic magnetic ordering phenomena in organic materials, in particular in pure graphite, has been one of the more exciting scientific activities working in the frontiers of physics, chemistry, and materials science. In this Letter we report on a novel chemical route leading to undoubtedly obtain macroscopic quantities of magnetic graphite. This material has a stable and strong ferromagnetic response even at room temperature where it can be attracted by a commercial magnet. We have obtained this magnetic graphite by a vapor reaction consisting of a controlled etching on the graphite structure. This behavior has been previously predicted and postulated to be associated to micro-structural characteristics breaking the continuity of the delocalized π-electron clouds of the graphitic material, thus allowing the existence of magnetic centers related to the topology.

[*] Author for correspondence and request of samples; e-mail address: amombru@fq.edu.uy

Carbon materials are attracting increasing attention due to the novelty of the associated physical properties and the potential applications in *high-tech* devices. The possibility to achieve outstanding properties in macroscopic carbon materials would open up a profusion of new striking applications. The production of bulk carbon magnetic material in a macroscopic amount, would have immediate impact through new and novel applications of this material in engineering, as well as in medicine and biology as a unique biocompatible magnetic material[1].

Magnetic properties induced by defects on graphite structures, such as pores, edges of the planes and topological defects, have been theoretically predicted. The possible coexistence of $sp^3$ and $sp^2$ bonds have been also invoked to predict this behavior[1]. Some reports have proved the existence of weak ferromagnetic-like magnetization loops in highly-oriented pyrolytic graphite (HOPG)[2,3]. Very recently two reports showed that the existence of ferromagnetism in pure carbon is unambiguously possible[4,5].

In this Letter we report on a novel and inexpensive chemical route consistent in a controlled etching on the graphite structure to obtain macroscopic amounts of bulk ferromagnetic pure graphite. This material has a strong magnetic response even at room temperature where it can be attracted by a commercial magnet and would be the experimental confirmation for the defect induced magnetism previously predicted.

The chemically modified graphite here reported was produced by a vapor phase *redox* reaction in closed nitrogen atmosphere ($N_2$, 1 atm.) with copper oxide (CuO). A few grams of both powders, CuO (Merck, analytical grade) and graphite (Fluka, CAS Number 7782-42-5, lot 426277/1, granularity < 0.1 mm) were placed at different alumina crucibles in a sealed atmosphere, inside a tube furnace (Fig. 1). The reaction took place at 1200ºC, during 24 hours. After the reaction was finished, the CuO was partially reduced to Cu(0) and, in the other container, the graphite showed a decrease in volume, specially in the side closest to the CuO crucible. Two clearly different regions could be observed: an upper layer, black and opaque, with amorphous aspect, and a lower layer, more crystalline than the original pure graphite. The material from the upper layer was the one showing the magnetic behavior, detectable up to room temperature. The separation of both materials was carefully achieved with the aid of a magnet.

This magnetic graphite was produced by the reaction of pristine graphite with controlled amounts of oxygen released from the decomposition of CuO at high temperature. This chemical attack created pores and stacking structures and increased the exposed edges of the graphene planes, producing a foamy-like graphite. The removed carbon may also recrystallize on the graphite crystallites surfaces at the same crucible, thus creating the so-called lower region. This vapor phase reaction took place in an inert gas environment. In this case both $N_2$ and Ar have been used with similar results, discarding the specific role of $N_2$ as a reagent or catalyst. The use of powder graphite as a reagent seems to be crucial because of the high reactivity to chemical attacks, due to the high exposed surface and the previous existence of defects. The reproducibility of the method and purity of the materials, with very special concern on the presence of metallic impurities, were regarded. Several samples have been prepared and all of them exhibited the magnetic behavior here described. Since the presence of any kind of ferromagnetic impurity must be avoided, we have carefully determined the chemical purity of the samples with AAS (atomic absorption spectroscopy) using a Shimadzu AA6800 spectrometer. The total content of iron in the samples was in the 40-60 *ppm* range. We have checked these results with XRF (x-ray fluorescence analysis) and EDS (energy dispersive spectroscopy), comparing the results obtained for the pristine and the modified graphite (Figs. 2 and 3, respectively). Although the wide background in EDS indicating the presence of carbon overlaps some transition metal peaks, the results are anyway evidence of the very low contamination (if any) caused by the procedure here presented. The results did not indicate any increase of metallic impurities with respect to the original pristine graphite. If these impurities would have been the cause of the magnetic effect here reported, both graphite samples, the modified and the pristine ones, should exhibit the same magnetic behavior, which clearly is not the case. No other ferromagnetic metals like nickel or cobalt were detected.

We have studied the obtained magnetic graphite samples by scanning electron microscopy (SEM), using a Jeol JSM 5900LV microscope. SEM micrographs have given enough evidences about the topological changes occurring at the micro-structural scale. Figure 3a. shows a micrograph from the pristine graphite powder used as reagent. Figure 3b. shows a graphite grain after chemical attack, exhibiting several pores distributed all over the sample. It also can be observed that a large dispersion of pore sizes can be achieved, with diameters ranging from a few nanometers to approximately 1 μm. Figure 3c. clearly shows a pore having about 1 μm diameter that goes through many graphene layers. In some regions where the etching process affected the boundaries of the graphene layers, other complex microstructures have been observed, like the stacking structure shown in Figure 3d.

We have magnetically characterized our treated graphite samples by performing magnetic measurements by using a MPMS-5T Quantum Design magnetometer. In Figure 4 we show the zero-field cooled (ZFC) magnetization versus temperature for an external magnetic field H = 1000 Oe. We have use the inverse susceptibility curve to determine the approximate value of the Curie temperature (see inset of Fig. 4).

Figure 5a exhibits the virgin hysteresis cycle for the m vs. H curves at 300 K, and Fig. 5b the same curve without the diamagnetic background. For this temperature, the saturation magnetic moment was very strong, 0.25 emu/g. To justify this value through the role of magnetic impurities by assuming that all these impurities behave as bulk ferromagnetic material -which most probably would not be the case-, it would be required about 1900 *ppm* of Fe. Since this value is much higher than the one determined as the total content of Fe in the studied samples, this justification could be ruled out. The coercive field was observed in the 300 K region was Hc = 350 Oe, and the remnant magnetization was 0.04 emu/g, which corresponds to 16 % of the saturation magnetic moment at 300 K. It is noteworthy that this room temperature ferromagnetic behavior of our treated graphite samples has also been verified through magnetic force microscopy experiments.

Preliminary results shows an exotic magnetic behavior of this material as a function of temperature, which will be reported elsewhere.

In summary, we have found a simple and inexpensive chemical route, based on a vapor phase reaction, to obtain bulk ferromagnetic graphite. According to theoretical studies[1], defects in the honeycomb structure of graphite could develop spontaneous magnetization due to the rise of a sharp asymmetric peak in the density of states at the Fermi level, which is required for a system of itinerant electrons to show ferromagnetism. The magnetic behavior here reported would be associated to microstructural features observed in the attacked sample that produce an inhomogeneous material with enhanced magnetism. The $sp^3$ and $sp^2$ bonds could play a role that should be investigated in the future.

The authors wish to thank PEDECIBA and CSIC (Uruguayan Organizations) and CNPq and FAPESP (Brazilian Organizations) for partial financial support. We gratefully acknowledge E. Longo, I. G. Gobato, A. Vercik, J. A. Chiquito, E. Marega, A. V. Narlikar, P. F. S. Moraes (*in memoriam*) and J. C. Ortega for valuable discussions and help in the initial steps of this work. We also thank the technical help of A. Márquez (SEM), A. Altamirano (AAS) A. Sixto and P. Noblía (XRF), and R. L. Almeida (magnetization measurements).

# Figure Captions

**Figure 1**. Sketch of the experimental reactor. (1) closed atmosphere, (2) CuO crucible, (3) graphite crucible, (4) tube furnace, (5) gas input, (6) vacuum.

**Figure 2**. EDS and XRF for (a) and (c) the pristine graphite and (b) and (d) the modified graphite. In the XRF figure the arrow shows the position of the FeK$\alpha$ peak. The intense peaks at higher energies correspond to the Ag target.

**Figure 3**. SEM images of the morphology of (**a)** the graphite used as reagent; (**b)** a region where many pores of different sizes can be seen; (**c)** one pore, 1 μm diameter –the propagation of the pore along the lamellar structure can be seen; **(d)** a lamellar stacking structure.

**Figure 4**. Zero-field cooled (ZFC) magnetization versus temperature curve, for an external magnetic field H=1000 Oe; the inset shows the inverse susceptibility curve where the arrow indicates the approximate value of the Curie temperature, $T \approx 323$ K.

**Figure 5**. Hysteresis curves, m vs. H, for the magnetic graphite material, for $T = 300$ K. (a) virgin curve; (b) after subtracting the diamagnetic background.

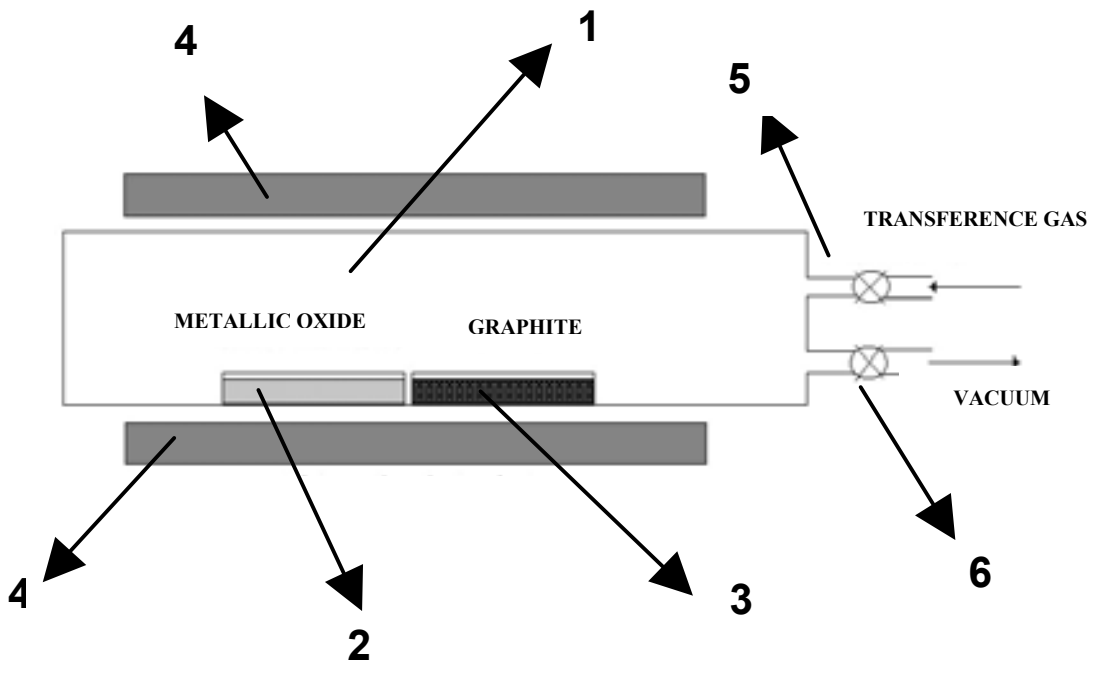

Figure 1 – H. Pardo *et al*.

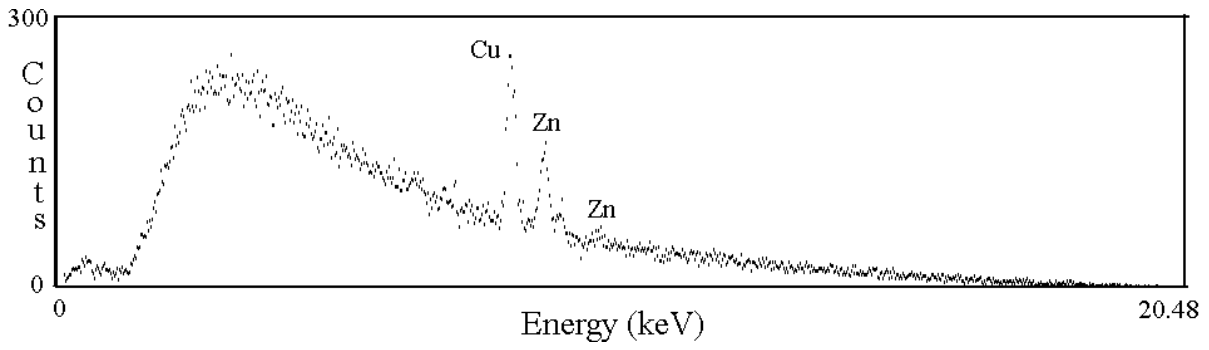

a)

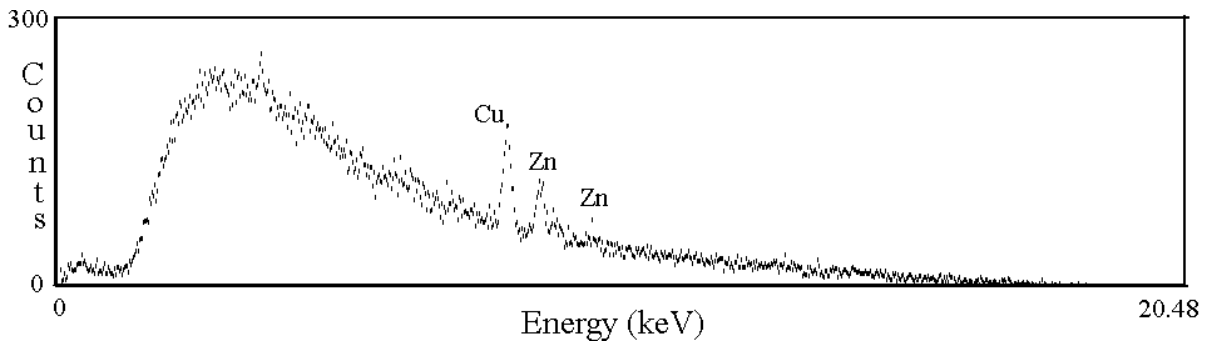

b)

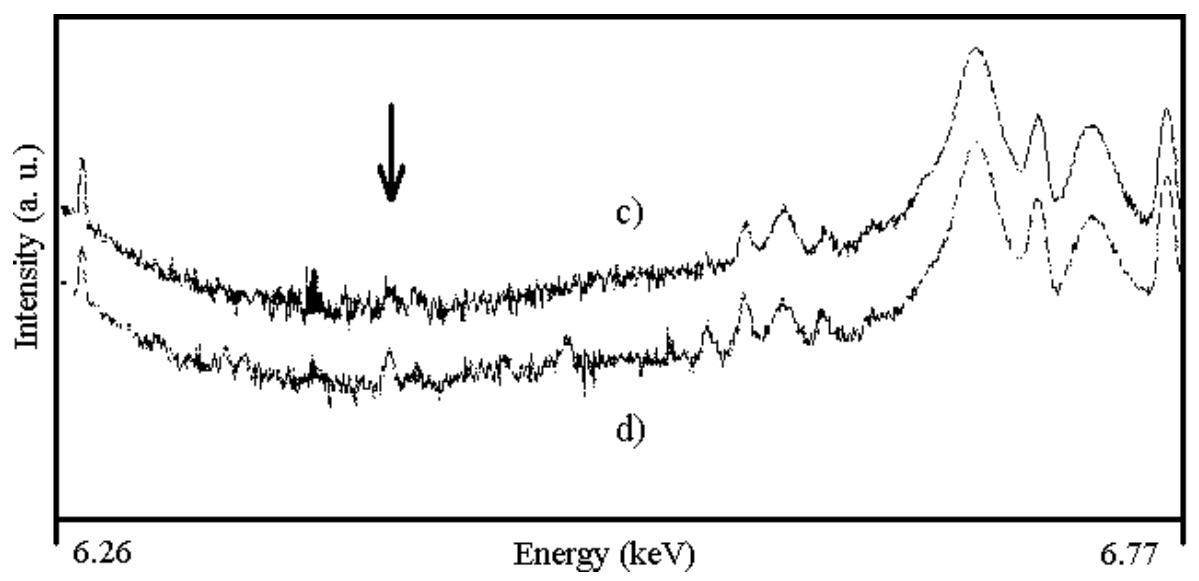

Figure 2 – H. Pardo *et al.*

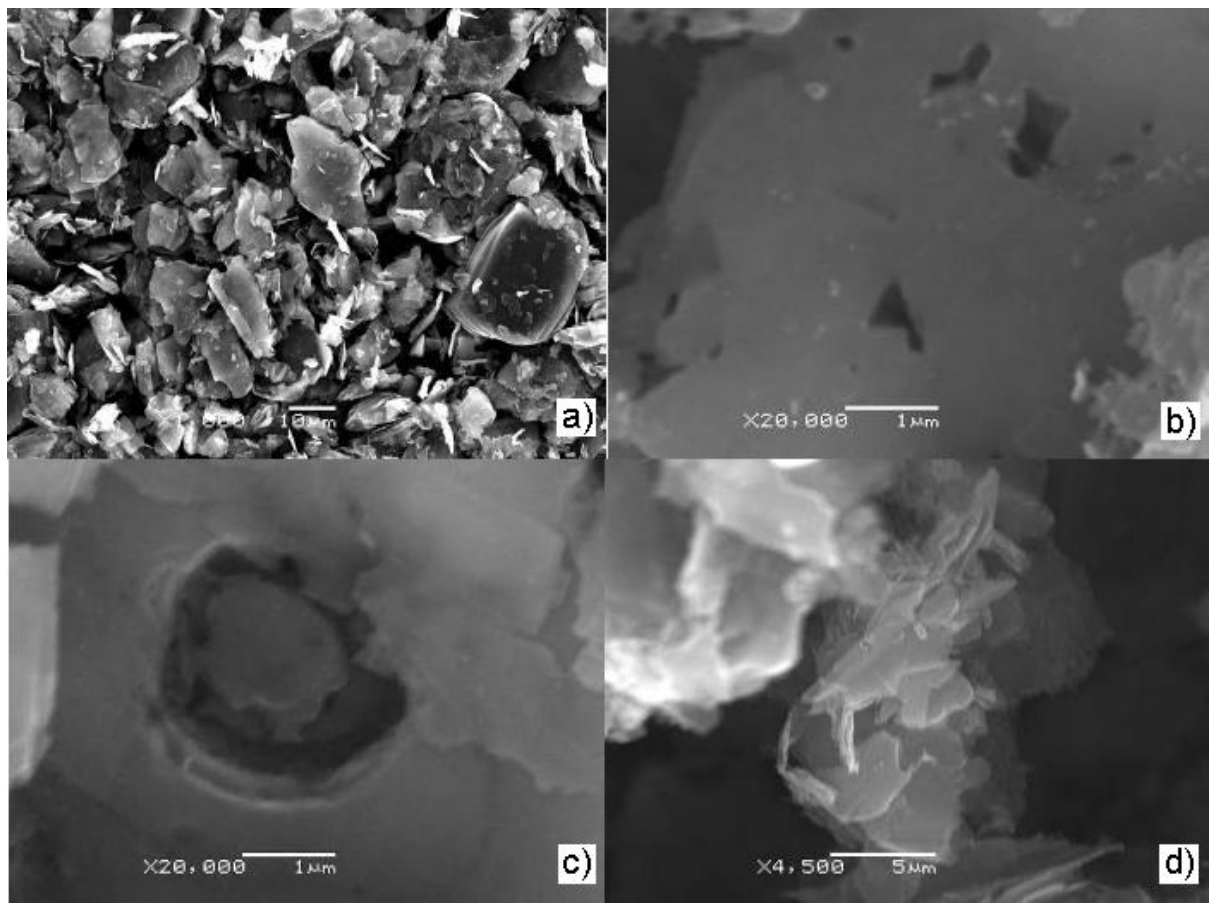

Figure 3 – H. Pardo *et al*.

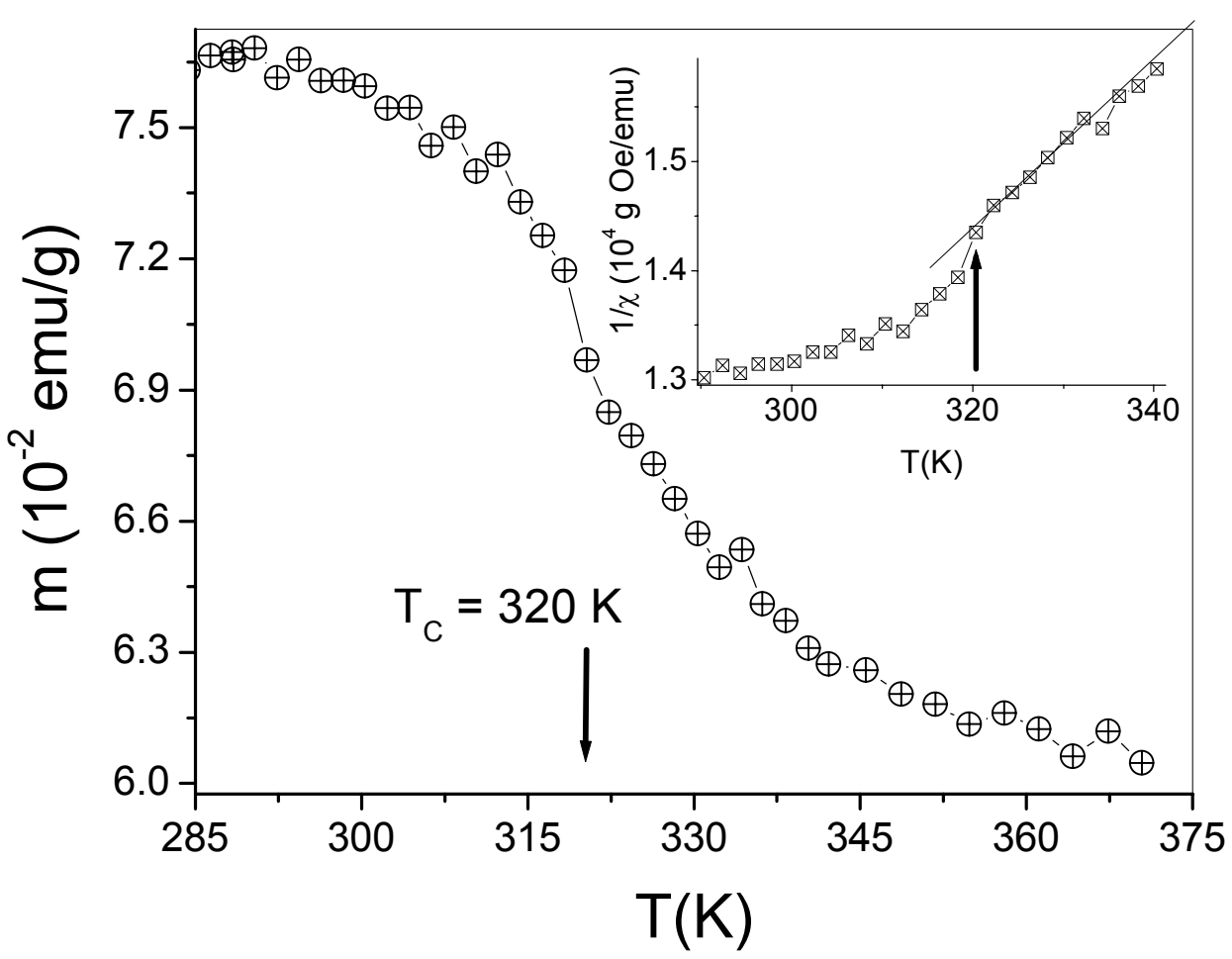

Figure 4 – H. Pardo *et al*.

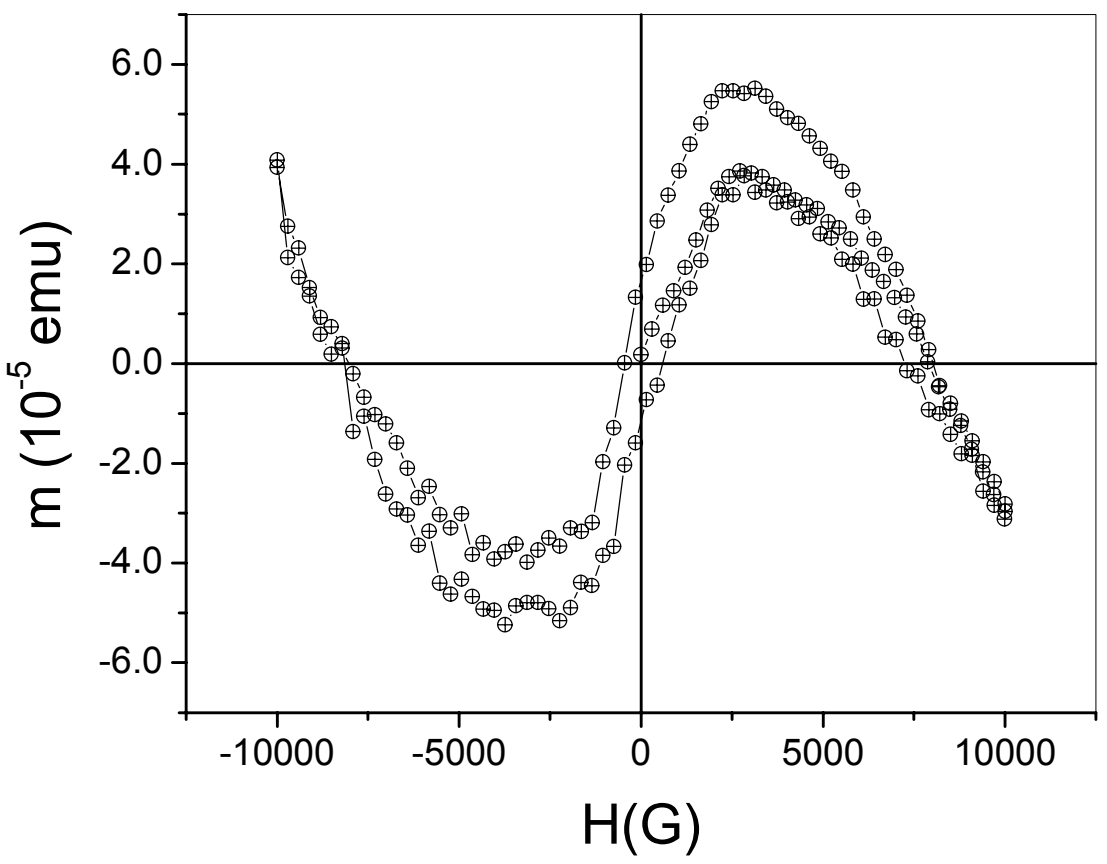

(a)

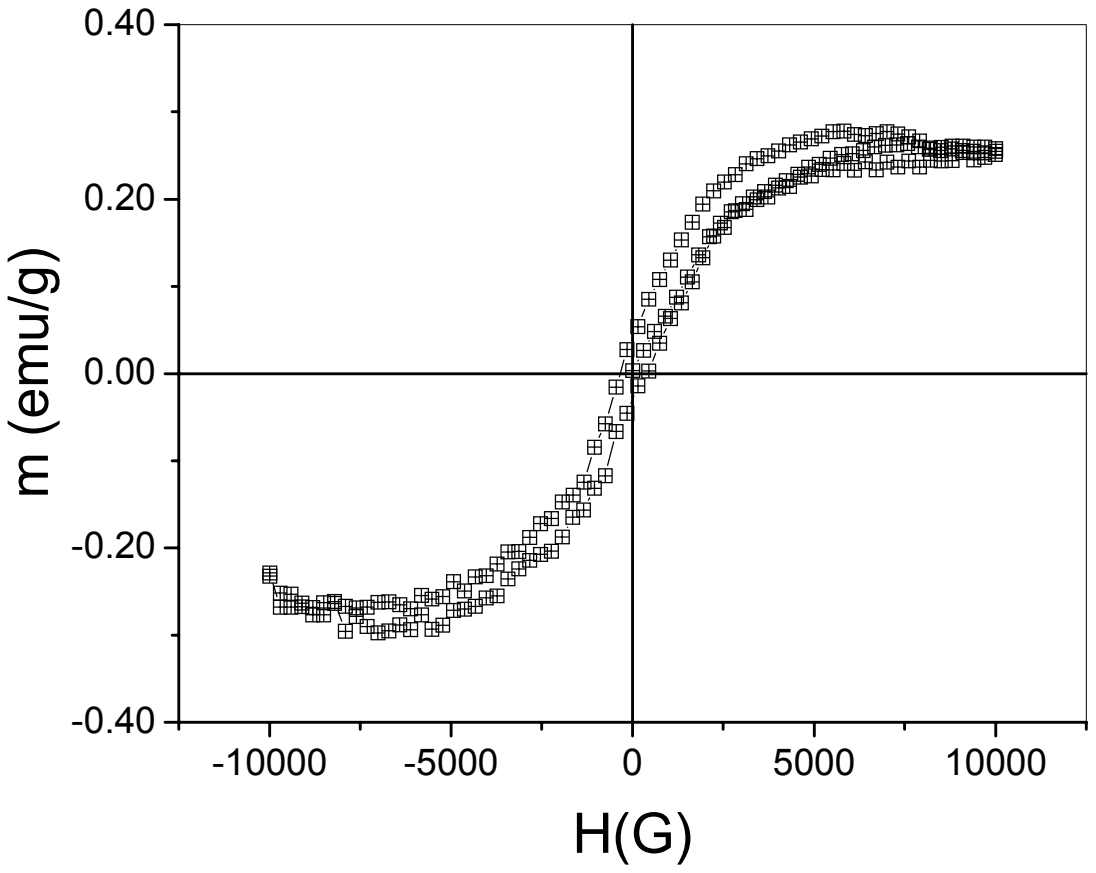

(b)

Figure 5 – H. Pardo *et al*.